\documentclass[pra,reprint]{revtex4-2}
\usepackage{graphicx}
\usepackage{enumitem}
\usepackage{amsmath}
\newcommand{\ket}[1]{|{#1}\rangle}
\newcommand{\bra}[1]{\langle{#1}|}

\usepackage{amsfonts}

\usepackage[usenames,dvipsnames]{xcolor}
\usepackage[colorlinks=true,citecolor=Cerulean,linkcolor=RubineRed,urlcolor=Cerulean]{hyperref}

\begin{document}

\title{Exact counterdiabatic driving for finite topological lattice models}

\author{Callum W. Duncan}
\email{callum.duncan@strath.ac.uk}
\affiliation{Department of Physics and SUPA, University of Strathclyde, Glasgow G4 0NG, United Kingdom}

\begin{abstract}
Adiabatic protocols are often employed in state preparation schemes but require the system to be driven by a slowly varying Hamiltonian so that transitions between instantaneous eigenstates are exponentially suppressed. Counterdiabatic driving is a technique to speed up adiabatic protocols by including additional terms calculated from the instantaneous eigenstates that counter diabatic excitations. However, this approach requires knowledge of the full eigenspectrum meaning that the exact analytical form of counterdiabatic driving is only known for a subset of problems, e.g., the harmonic oscillator and transverse field Ising model. We extend this subset of problems to include the general family of one-dimensional non-interacting lattice models with open boundary conditions and arbitrary on-site potential, tunnelling terms, and lattice size. We formulate this approach for all states of lattice models, including bound and in-gap states which appear, e.g., in topological insulators. We also derive targeted counterdiabatic driving terms which are tailored to enforce the dynamical state to remain in a specific state. As an example, we consider state transfer using the topological edge states of the Su-Schrieffer-Heeger model. The derived analytical counterdiabatic driving Hamiltonian can be utilised to inform control protocols in many-body lattice models or to probe the non-equilibrium properties of lattice models. 
\end{abstract}
\pacs{}

\maketitle

\section{Introduction}\label{sec:intro}

The adiabatic approximation is utilised to enable a variety of aspects of quantum science. This includes state preparation, the paradigm of adiabatic quantum computing \cite{aharonov2008adiabatic,albash2018adiabatic}, and coherent quantum annealing \cite{somma2012quantum,rajak2023quantum}. It states that if we initiate a system in an eigenstate of its Hamiltonian, then the dynamical solution to the Schr\"odinger equation as we change a parameter in said Hamiltonian, will approximately remain in the corresponding instantaneous eigenstate up to a phase factor \cite{born1928beweis,kato1950adiabatic}. This approximation relies upon the parameter changing slowly, as it is natural in such a scenario to describe the dynamical state entirely in the adiabatic basis of the instantaneous eigenstates of the changing Hamiltonian. If the state remains non-degenerate and the gap between it and all other states is large with respect to the inverse of the time taken to traverse the energy landscape, then the adiabatic approximation is valid \cite{jansen2007bounds}. This makes the adiabatic approximation difficult to realise if the parameter change includes the crossing of a phase transition or, as we will see below, the closing of a band gap. In any finite time protocol, there will always be a non-zero probability of transitioning to other states, and this is exasperated in most experimental settings due to the limited time allocated to an adiabatic protocol as well as losses, heating, noise, and dissipation. Therefore, speeding up adiabatic protocols, or finding alternative state preparation schemes, is crucial.

There are many approaches to the particular problem of speeding up an adiabatic protocol, including numerical optimal control \cite{glaser2015training,werschnik2007quantum} and shortcuts to adiabaticity \cite{odelin2019shortcuts,torrontegui2013shortcuts}. In this work, we will extend the analytical approach of Counterdiabatic Driving (CD) to include the general family of Hamiltonians of non-interacting particles in 1D lattices. CD was first introduced by Demirplak and Rice \cite{demirplak2003,demirplak2005} in quantum chemistry before being independently introduced as transitionless driving by Berry \cite{berry_transitionless_2009}. CD adds control terms to the dynamical Hamiltonian such that the adiabatic approximation is enforced as the solution of the dynamical Schr\"odinger equation for all timescales. It is found that by solving for the instantaneous eigenstates at all times, such a control term can be constructed and we will go through this in detail in Sec.~\ref{sec:CD}. Due to CD's reliance on knowledge of the instantaneous eigenstates, its exact form is only known for a very limited set of problems, e.g., harmonic oscillators \cite{campo2013shortcuts} and the integrable transverse Ising model \cite{campo2012assisted,damski2014counterdiabatic}, to which we will add non-interacting lattice models.

For a broader context, we note that approaches have been developed to give approximate CD terms for complex settings where it is not possible to construct the exact CD, with the most successful to date being that of local, or variational, CD \cite{sels2017minimizing,claeys2019floquet,kolodrubetz2017geometry}. This approach relies on recasting the CD terms into a description in terms of the adiabatic gauge potential, which encodes all of the diabatic transitions that one needs to counter and is equivalent to the CD term we will define in Sec.~\ref{sec:CD}. An approximation to the adiabatic gauge potential is then obtained through a variational minimisation procedure. The case of non-interacting particles in 1D lattices was considered as an example in the original proposal for local CD \cite{sels2017minimizing} and an approximate adiabatic gauge potential was obtained. The approach of local CD has been further developed to include its combination with numerical optimal control to improve annealing protocols \cite{cepaite2023counterdiabatic} and to inform the structure of terms utilised in reinforcement learning for optimal control \cite{yao2021reinforcement}. Local CD has been implemented experimentally for adiabatic state transfer in a one-dimensional lattice \cite{meier2020counterdiabatic} and has been extended to many-body lattice models \cite{xie2022variational}. Recently, numerical approaches to derive the adiabatic gauge potential in general many-body settings have been introduced either utilising Krylov subspace methods~\cite{takahashi2024shortcuts} or the Lie-algebra of the expansion~\cite{lawrence2023numerical}.

While the methods presented in this work can be applied to any general non-interacting 1D lattice model, we will focus on the particularly interesting case of time-reversal-symmetric topological insulators which can host edge modes in the band gaps of their spectrum \cite{hasan2010collouium,qi2011topological,rachel2018interacting,asboth2016short}. While direct brute-force numerical diagonalisation of the time-independent Schr\"odinger equation is always an option, analytical approaches have been developed to characterise the topological states in cases of semi-infinite commensurate lattices \cite{hatsugai1993chern,hugel2014chiral} through the extension of Bloch's theorem. It is also possible to obtain the topological edge states from the bound states of scattering matrix approaches \cite{fulga2012scattering,arkinstall2017topological}. We will utilise the approach of Ref.~\cite{duncan2018exact}, which allows for all states to be obtained, both within energy bands and in the gaps between them, for general 1D non-interacting topological lattice models with open boundary conditions. We will outline this approach in Sec.~\ref{sec:states} which will enable us to write the general analytical form of CD for non-interacting lattice models in Sec.~\ref{sec:CDlat}. We will then consider as an example the topological Su-Schrieffer-Heeger (SSH) model which has been used as a toy model for the control of topologically protected state transfer protocols \cite{qi2020engineering,mei2018robust,longhi2019landau,palaiodimopoulos2021fast,angelis2020fast,qi2020controllable,wang2022arbitrary}. We will show both the form of the analytically obtained CD for the SSH model and consider the properties of the modified dynamical Hamiltonian which enforces adiabaticity.  

\section{Counterdiabatic driving}\label{sec:CD}

CD enforces the adiabatic approximation to be the dynamical solution to Schr\"odinger's equation through the addition of control terms \cite{demirplak2003,demirplak2005,berry_transitionless_2009}. For an arbitrary time-dependent Hamiltonian $H(\lambda(t))$, where $\lambda(t)$ is the changing parameter, the instantaneous eigenstates,  $\ket{\psi_n(\lambda(t))}$, and energies, $E_n \left(\lambda(t)\right)$, are given by
\begin{equation}
    H\left(\lambda(t)\right) \ket{\psi_n\left(\lambda(t)\right)} = E_n \left(\lambda(t)\right) \ket{\psi_n\left(\lambda(t)\right)} \ ,
\end{equation}
with $n$ being the quantum number of the eigenstates. The adiabatic approximation then gives the solution of the dynamical Schr\"odinger's equation to be
\begin{equation}\label{eq:adiabatic}
    \ket{\Psi_n\left(\lambda(t)\right)} = \exp \left( - \frac{i}{\hbar} \int_{t_0}^{t_f} dt E_n \left(\lambda(t)\right) \right) \ket{\psi_n\left(\lambda(t)\right)} \ ,
\end{equation}
where we have excluded the Berry phase which would give the geometric phase in a cyclic protocol \cite{berry1984quantal}. From here we will drop the explicit time dependence of the parameter $\lambda$ and work in units of $\hbar=1$. We also note that the arguments of this section, and the rest of this work, can be easily extended to the case of a set of changing parameters.

To enforce Eq.~\eqref{eq:adiabatic} to be the solution to the dynamical Schr\"odinger equation in arbitrary driving times, we are required to add to the original Hamiltonian the CD term of \cite{berry_transitionless_2009}
\begin{equation}\label{eq:CD}
    H_{\mathrm{CD}} = i \sum_n \dot{\lambda} \ket{\partial_\lambda \psi_n(\lambda)} \bra{\psi_n(\lambda)}\ ,
\end{equation}
where $ \dot{\lambda}$ represents the derivative of the parameter $\lambda$ with respect to time. In order to construct the exact CD term we therefore need to be able to solve the instantaneous Schr\"odinger equation for all points along the path dictated by $\lambda$.

\section{Exact states of non-interacting lattice models}\label{sec:states}

We will briefly introduce the known solutions to the general non-interacting problem on a lattice described by the Hermitian Hamiltonian
\begin{equation}
    H = \sum_{x=x_0+1}^{L-1} \left( - J_x b^\dagger_x b_{x+1} - J_x^* b^\dagger_{x+1} b_x + \mu_x n_x\right) \ ,
\end{equation}
with the lattice being $L$ sites labelled by $x$ which takes consecutively increasing integer values between $x_0+1$ and $L-1$, $b^\dagger_x$ ($b_x$) being the creation (annihilation) operator of a particle on the site at position $x$, $n_x$ the number operator on the site at $x$, $J_x$ the tunnelling strength between site $x$ and $x+1$, and $\mu_x$ the on-site potential for site $x$. We will consider open boundary conditions and crystalline models where the system has a finite unit cell and thus a periodicity, which we label as $\tau$, allowing us to simplify the problem via Bloch's theorem. However, the techniques outlined here can be applied in other one-dimensional lattice cases. 

The non-interacting states can be written as
\begin{equation}\label{eq:stategen}
    \ket{\psi_\alpha} = \sum_{x=x_0+1}^{L-1} \psi_\alpha \left( x \right) b^\dagger_x \ket{0} \ ,
\end{equation}
with $\ket{0}$ being the state with no particles, and $\psi_\alpha(x)$ being the coefficients of the state in each site $x$. We will refer to $\psi_\alpha(x)$ as the wave function as it fully describes the quantum state. The general wave function for both states within a band or in a band gap, e.g. topological edge states \cite{hasan2010collouium,asboth2016short} or Shockley-type bound states \cite{shockley1939on}, can be written as \cite{duncan2018exact}
\begin{equation}\label{eq:state}
    \psi_\alpha(x) = N \left[ \phi_+ (x) \alpha^x - \frac{\phi_+ (L)}{\phi_- (L)} \phi_- (x) \alpha^{2L-x} \right] \ ,
\end{equation}
with $0<|\alpha|<1$ being the parameter which fully characterises the individual states and $N$ a normalisation factor. The Bloch functions $\phi_{+(-)} (x)$ correspond to those associated with $\alpha$ ($\alpha^{-1}$). These Bloch functions can be obtained using Bloch's theorem, i.e. by considering the local Schr\"odinger equation for each site in the unit cell and individually taking an ansatz of either $\phi_+(x)\alpha^x$ or $\phi_-(x)\alpha^{-x}$ then using $\phi_{\pm}(x) = \phi_\pm(x+\tau)$ so that the Bloch functions are obtained after solving $\tau-1$ linear coupled equations. Note, in the course of solving for the Bloch functions we will also obtain the analytical form of the energy spectrum, $E(\alpha)$. The normalisation factor for each state can be calculated using the Bloch functions and the quantised values of $\alpha$ as
\begin{equation}
    N = \left[ \sum_{x=x_0+1}^{L-1} \left| \phi_+ (x) \alpha^x - \frac{\phi_+ (L)}{\phi_- (L)} \phi_- (x) \alpha^{2L-x} \right|^2 \right]^{-1/2} \ .
\end{equation}

The quantisation of $\alpha$ to give a finite number of states is dependent on the boundary condition and the form of the Hamiltonian. We can solve for $\alpha$ in two ways, either by taking the general quantisation condition given by the boundary conditions, $\psi(x_0)=\psi(L)=0$, to obtain
\begin{equation}\label{eq:quant}
    \alpha^{2(L - x_0)} = \frac{\phi_+(x_0)\phi_-(L)}{\phi_+(L)\phi_-(x_0)} \ ,
\end{equation}
or by considering the Schr\"odinger equation at a single site, we will consider it at site $x_0+1$ and use $\psi(x_0)=0$ to obtain
\begin{equation}\label{eq:localSE}
    E(\alpha) = J_{x_0+1} \frac{\psi(x_0+2)}{\psi(x_0+1)} + \mu_{x_0+1} \ .
\end{equation}
When solving for a bound state in the system we will need to solve the local Schr\"odinger equation given by Eq.~\eqref{eq:localSE}, as the presence of a bound state does not solely rely on the boundary conditions which is what leads us to the quantisation condition of Eq.~\eqref{eq:quant}.

States contained within energy bands and not within band gaps are described in terms of plane waves with real quasimomentum $k$. This means that a number of the solutions obtained from the above equations will be of the form $\alpha=e^{i k}$ and if no in-gap states are present then all the solutions will be of this form. It is worth noting that for a commensurate system, $\phi_\pm(x_0) = \phi_\pm(L)$, we can solve for the quasimomenta for any Hamiltonian without even defining the Hamiltonian, as the quantisation condition of Eq.~\eqref{eq:quant} in this case is $e^{2i(L-x_0)k}=1$, which has known solutions of
\begin{equation}
    k = \frac{\pi n}{L-x_0} \ ,
\end{equation}
with $n \in \mathbb{Z}$ being the quantum number that characterises the different eigenstates. The fixed quasimomenta of commensurate lattices will allow us to simplify the CD terms for them substantially, as we will outline below.

\section{Exact counterdiabatic driving in lattices}\label{sec:CDlat}

To write the counderdiabatic terms of non-interacting lattice models we first need to outline the construction of the operator summed over in Eq.~\eqref{eq:CD}. The eigenstates of general 1D lattice models are given by Eq.~\eqref{eq:stategen} with the wave function $\psi_\alpha(x)$ given in Eq.~\eqref{eq:state}. Using these, we can then take the derivative w.r.t. $\lambda$ to obtain 
\begin{equation}
    \partial_\lambda \psi_\alpha (x) = N \left[ \phi_+(x) \alpha^x A_\alpha(x) - \frac{\phi_+(L)}{\phi_-(L)} \phi_-(x) \alpha^{2L-x} B_\alpha(x) \right] \ ,
\end{equation}
where we have defined 
\begin{equation}
    A_\alpha(x) = \frac{\partial_\lambda N}{N} + \frac{\partial_\lambda \phi_+(x)}{\phi_+(x)} + x\frac{\partial_\lambda \alpha}{\alpha} \ ,
\end{equation}
and
\begin{equation}
\begin{aligned}
    B_\alpha(x) & = & \frac{\partial_\lambda N}{N} + \frac{\partial_\lambda \phi_+(L)}{\phi_+(L)} -  \frac{\partial_\lambda \phi_-(L)}{\phi_-(L)} \\ & & + \frac{\partial_\lambda \phi_-(x)}{\phi_-(x)} + \left( 2L - x \right) \frac{\partial_\lambda \alpha}{\alpha}
    \end{aligned} \ .
\end{equation}

We will now outline how each term for $A_\alpha(x)$ and $B_\alpha(x)$ can be analytically obtained. First, the terms which are derivatives of Bloch functions can be obtained after these are solved for in a given model. We will discuss this for an the example in Sec.~\ref{sec:SSH} and can not go further here without knowing the form of the Bloch functions. 

Next, we consider the term $\partial_\lambda N$, which we can write as
\begin{equation}
    \partial_\lambda N = -\frac{1}{2} N^3 \sum_{x=x_0+1}^{L-1} \left[ \Tilde{\psi}^*_\alpha(x) \partial_\lambda \Tilde{\psi}_\alpha(x) + \Tilde{\psi}_\alpha(x) \partial_\lambda \Tilde{\psi}^*_\alpha(x) \right] \ ,
\end{equation}
with $\Tilde{\psi}_\alpha(x) = \psi_\alpha(x)/N$, i.e., the wave function without normalisation. The derivative of the unnormalised wave functions can be written out as
\begin{equation}
    \partial_\lambda \Tilde{\psi}_\alpha(x) = \phi_+(x) \alpha^x \Tilde{A}_\alpha(x) - \frac{\phi_+(L)}{\phi_-(L)} \phi_-(x) \alpha^{2L-x} \Tilde{B}_\alpha(x) \ ,
\end{equation}
with
\begin{equation}
    \Tilde{A}_\alpha(x) = A_\alpha(x) - \frac{\partial_\lambda N}{N} \ ,
\end{equation}
and
\begin{equation}
    \Tilde{B}_\alpha(x) = B_\alpha(x) - \frac{\partial_\lambda N}{N} \ .
\end{equation}

Finally, we are left with the term proportional to $\partial_\lambda \alpha$. For commensurate systems where the quasimomentum of the bulk states is well-defined by $k=n\pi/(L-x_0)$, see Sec.~\ref{sec:states}, the term $\partial_\lambda \alpha$ will be zero for all states within the bulk energy bands. This is because for these states $\alpha=e^{ik}$ and the quasimomentum is entirely defined from the boundary conditions. Note, in this case, the $\partial_\lambda N$ is also zero, and we can simplify the expressions for $A_\alpha(x)$ and $B_\alpha(x)$ considerably to only rely on the Bloch functions. This still leaves us with needing to know $\partial_\lambda \alpha$ for in-gap states or for bulk states in incommensurate lattices. We can obtain this by differentiating the local Schr\"odinger equation of Eq.~\eqref{eq:localSE} w.r.t. $\lambda$. Once we know $\alpha$ we can then simply substitute this in and rearrange this new differentiated local Schr\"odinger equation to analytically obtain $\partial_\lambda \alpha$.

\subsection{All states}

We can now write the form of the additional CD term to counter transitions between all instantaneous eigenstates as
\begin{equation}
    H_\mathrm{CD} = i  \sum_\alpha \sum_{x,x^\prime=x_0+1}^{L-1} \theta_\alpha(x,x^\prime) b^\dagger_x b_{x^\prime} \ ,
\end{equation}
where the summation runs over both $x$ and $x^\prime$ such that the CD term couples all sites in the chain. The strength of this coupling is described by the function
\begin{equation}\label{eq:CDall}
\begin{aligned}
    \theta_\alpha(x,x^\prime) = \: & \left| N \right|^2 \psi^*_\alpha(x^\prime) \Bigg[\phi_+(x) \alpha^x A_\alpha(x)  \\ & - \frac{\phi_+(L)}{\phi_-(L)} \phi_-(x) \alpha^{2L-x} B_\alpha(x)\Bigg] \ .
\end{aligned}
\end{equation}

\subsection{Targeted states}

In certain situations, e.g., in ground state coherent quantum annealing, we will only be interested in staying in one single state of the system and it is desirable to only counter the diabatic terms out of this target state. We propose that this targeted CD can be realised by the addition of terms of the form
\begin{equation}\label{eq:CDparticularH}
    H_\mathrm{CD}^\alpha = i \ket{\partial_\lambda \psi_\alpha(x)} \bra{\psi_\alpha (x)} - i \ket{\psi_\alpha (x)} \bra{\partial_\lambda \psi_\alpha(x)} \ .
\end{equation}
Similar forms for CD of particular states have been considered previously \cite{demirplak2008consistency,zheng2016cost,campbell2023quantum}, especially when trying to reduce the energetic overhead for implementation. We can then write this for non-interacting lattice models as
\begin{equation}\label{eq:CDparticular}
    H_\mathrm{CD}^\alpha = i \sum_{x,x^\prime=x_0+1}^{L-1} \left[ \theta_\alpha(x,x^\prime) b^\dagger_x b_{x^\prime} - \theta^*_\alpha(x,x^\prime) b^\dagger_{x^\prime} b_x \right] \ ,
\end{equation}
with $\alpha$ being that of the targeted state, e.g., the instantaneous ground state for coherent quantum annealing. We will see in the example of Sec.~\ref{sec:SSH} that the targeted CD term can result in a simplification of the control terms that need to be implemented.

\section{State transfer in the topological Su-Schrieffer-Heeger model}\label{sec:SSH}

\subsection{The Su-Schrieffer-Heeger model}

\begin{figure}[t]
	\centering
	\includegraphics[width=0.98\linewidth]{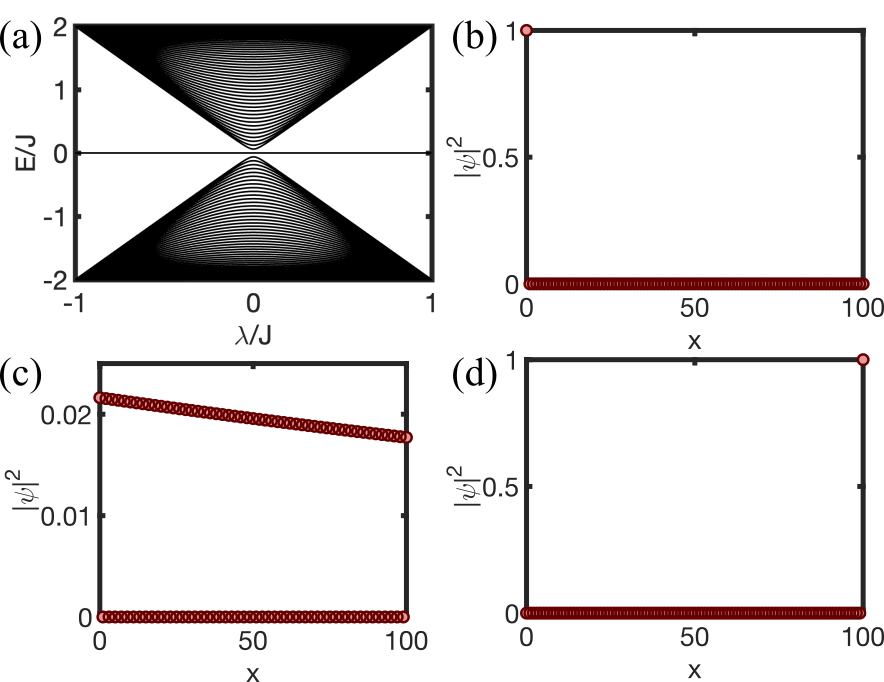}
	\caption{Spectrum and edge states of the SSH model given by Hamiltonian~\eqref{eq:SSHH} using the analytical energy and states for a system of $101$ sites. (a) the spectrum of all states for each individual $\lambda/J$ as given by Eq.~\eqref{eq:SSHE} with the states in the bulk having $\alpha=e^{ik}$ with a quasimomentum $k=\pi n/102$, with $n\in \mathbb{Z}$ starting at one and monotonically increasing for each state, and the in-gap edge state being described by an $\alpha$ obtained by solving the Schr\"odinger equation at the first site. We show the probability density for (b) the in-gap state at $\lambda/J=0.999$ with $\alpha=0.0224e^{i\pi/2}$ and occupation on only even sites, (c) the in-gap state at $\lambda/J=10^{-3}$ with $\alpha=0.999e^{i\pi/2}$, and (d) the in-gap state at $\lambda/J=-0.999$ with $\alpha=0.0224e^{i\pi/2}$.}
	\label{fig:SSHExample}
\end{figure}

As an example, we will consider adiabatic state transfer through the topological edge states of a one-dimensional model, the Su-Schrieffer-Heeger (SSH) model. The SSH model was first introduced as a model of polyacetylene \cite{su1979solitons} and has topological edge states with a corresponding non-trivial Zak's phase \cite{zak1989berry}. This model has been realised in ultracold atoms \cite{meier2016observation}, the energy levels of a Rydberg atom \cite{kanungo2022realizing}, photonic lattices \cite{malkova2009observation,st2017lasing,xia2021topological}, and acoustic waveguides \cite{coutant2021acoustic}.  There has been particular interest in studying topologically protected state transfer in the SSH model and the optimisation of such a transfer, including removing the requirement for adiabatic evolution \cite{qi2020engineering,mei2018robust,longhi2019landau,palaiodimopoulos2021fast,angelis2020fast,qi2020controllable,wang2022arbitrary}.

The SSH model has no on-site potential and has nearest-neighbour tunnelling which alternate in strength, meaning we can write the Hamiltonian as
\begin{equation}\label{eq:SSHH}
    H(\lambda) = \sum_{x=x_0+1}^{L-1} \left[ \left( 1 - \lambda (-1)^x \right) b^\dagger_x b_{x+1} + \mathrm{H.c.} \right] \ ,
\end{equation}
with $2 \lambda$ being the difference in strength between the alternating tunnellings.  We will consider a commensurate finite system, i.e., $\phi_\pm(L) = \phi_\pm(x_0)$. For the commensurate lattice, we can perform a state transfer between the two edges of the system by initialising the system on the $x_0+1$ edge with a large $\lambda_0$ then driving the system towards $-\lambda_0$. If the sweep through $\lambda$ is performed adiabatically, the final result will be the perfect transfer of the state to the $L-1$ edge. This state transfer protocol is illustrated in Fig.~\ref{fig:SSHExample}, including examples of the edge state for different $\lambda$.

Using the approach outlined in Sec.~\ref{sec:states}, the spectrum 
\begin{equation}\label{eq:SSHE}
    E_{s,\alpha} = (-1)^s \sqrt{\left( \frac{1+\alpha^2}{\alpha} \right)^2 - \lambda^2 \left( \frac{\alpha^2-1}{\alpha} \right)^2} \ ,
\end{equation}
with $s=0,1$ labelling the two bands, and Bloch functions
\begin{equation}\label{eq:BlochSSH}
    \phi_+(x) = \begin{pmatrix}
    1 \\
    \frac{(1+\alpha^2)-\lambda(1-\alpha^2)}{E_{s,\alpha}\alpha}
    \end{pmatrix} \ ,
\end{equation}
can be found. To obtain $\phi_-(x)$ we implement the transformation $\alpha\rightarrow 1/\alpha$ to Eq.~\eqref{eq:BlochSSH}. Note that as would be expected, $E_{s,\alpha} \equiv E_{s,1/\alpha}$.

\begin{figure*}[t]
	\centering
	\includegraphics[width=0.98\linewidth]{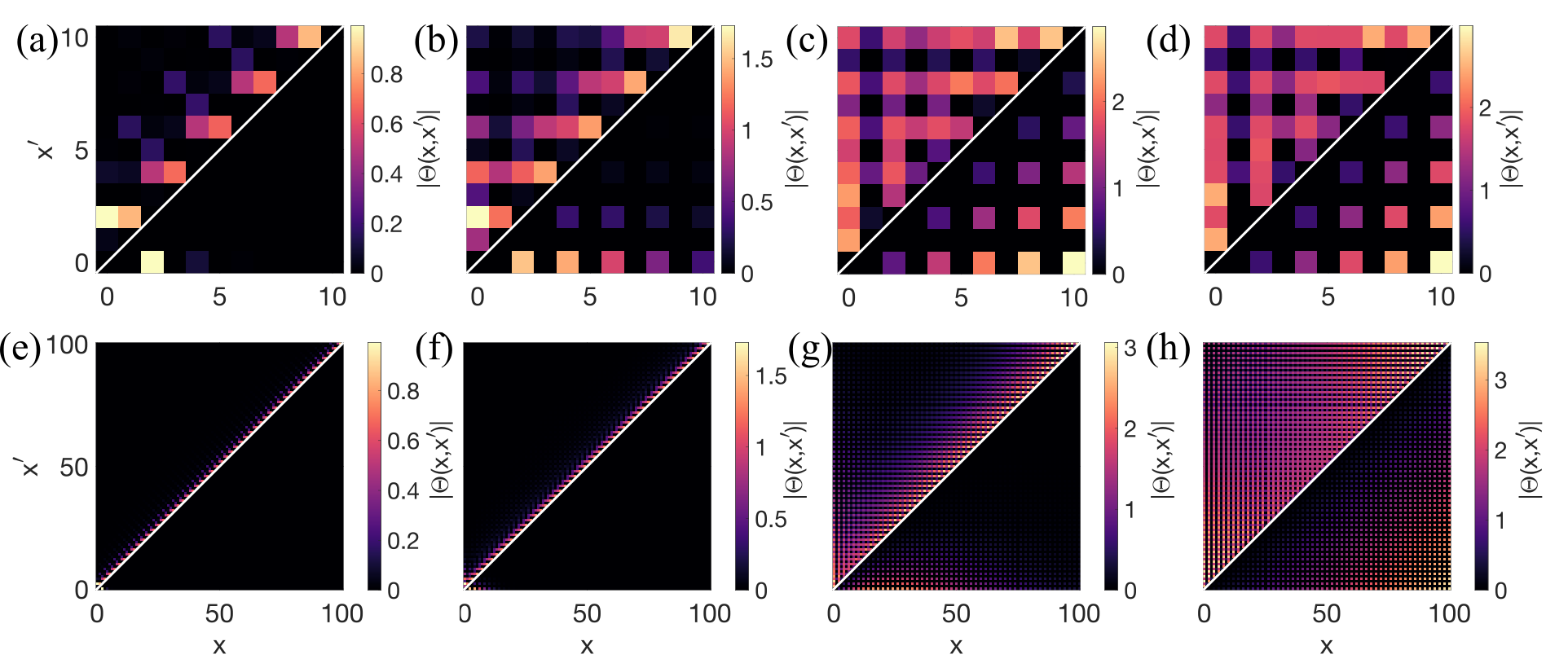}
	\caption{Examples of the form of the CD terms through plots of the absolute value in each element of the matrix of the CD term for the same driving as considered in Fig.~\ref{fig:CDspectra}. The CD terms are Hermitian and each image shows the form of the full CD term $H_\mathrm{CD}$ in the upper triangle and the targeted CD term $H_\mathrm{CD}^\alpha$ using only the in-gap state in the lower triangle. Note, that diagonal terms are trivially zero for CD and we have separated the triangles with a white line. We consider two different system sizes (a-d) $11$ and (e-h) $101$ sites as well as a number of different points along the diving with (a,e) $\lambda/J=0.9$, (b,f) $\lambda/J=0.36$, (c,g) $\lambda/J=4.6\times 10^{-2}$, and (d,h) $\lambda/J=10^{-3}$.}
	\label{fig:CDSSH}
\end{figure*}

The spectrum for a commensurate system with $\phi_\pm(L) = \phi_\pm(x_0)$ as a function of $\lambda$ is shown in Fig.~\ref{fig:SSHExample}(a) for $x_0=-1$ and $L=101$. We already know all the solutions in the two bulk bands, as the quantisation condition for commensurate systems enforces $k=\pi n/(L-x_0)$. For finite $\lambda$ there are two bands, as there are two sites in the unit cell. This means we only need to solve for the missing state which is in the band gap, this can be done by solving the local Schr\"odinger equation of Eq.~\eqref{eq:localSE} for $\alpha$. Note, this will give us all of the bulk solutions as well, but the problem can be simplified as we know the quasimomentum must be that of the missing state, $k=\pi/2$ \cite{duncan2018exact}. This means we can take $\alpha = a e^{i \pi/2}$ where $a \in \mathbb{R}$ and solve Eq.~\eqref{eq:localSE} for $a$, giving us a single solution, that of the in-gap edge state. Note, at $\lambda=0$ we return to a single band model and $a=1$. With positive $\lambda$ the edge state will be bound to the $x_0$ site boundary and for negative $\lambda$ to the $L$ site boundary, with examples shown in Fig.~\ref{fig:SSHExample}(b) and (d). Therefore, if we follow the in-gap state adiabatically from positive to negative $\lambda$ in a dynamical protocol, we will transfer a particle from one edge to the other.

However, there is an issue in realising this adiabatic protocol, the gap closes at $\lambda=0$ as can be seen in Fig.~\ref{fig:SSHExample}(a) and from the fact that $a=1$ at this point, i.e., the solution is that of a bulk state in a band. To be adiabatic one would need to drive slowly through this region, with a rate of change of the parameter being inversely proportional to the gap between the state we desire to remain in and the nearest other eigenstate. As we approach $\lambda=0$, the edge state begins to stretch further into the system, as is shown for $\lambda/J = 10^{-3}$ in Fig.~\ref{fig:SSHExample}(c). In this example, we will use the approaches outlined in Sec.~\ref{sec:CDlat} to study the diabatic terms that arise from the closing of this gap and what is needed to implement perfect state transfer in arbitrary time. 

For the SSH model, we can confirm the gap goes to zero for large $L$ (and fixed $x_0$) and obtain its scaling. As we are working with a commensurate system, $\phi_\pm(L) = \phi_\pm(x_0)$, we know the quasimomenta for all of the bulk states and we know that the edge state has $k=\pi/\tau$. Using the fact that the edge state will have zero energy and that the spectrum is symmetric, we can calculate the gap simply by looking at the energy of the previous state which will have quantum number $n=(L-1)/2$ for the commensurate case and therefore $k=\pi (L-1)/2(L-x_0)$. Fixing $x_0=-1$ we find the gap is given by
\begin{equation}\label{eq:SSHgapclose}
    \Delta E(\lambda) = \sqrt{2} \sqrt{1 + \lambda^2 + (1-\lambda^2) \cos \left(\frac{(L-1)\pi}{L+1} \right)} \ .
\end{equation}
Which has $\lim_{L \rightarrow \infty} \Delta E (\lambda) = 2\lambda$, as $\lim_{L \rightarrow \infty} \frac{L-1}{L+1} = 1$.

\subsection{Counterdiabatic driving}

In constructing the CD terms, we know that the derivatives of the Bloch functions w.r.t. the varying parameter play a vital role. For the SSH model, we can find these by differentiating Eq.~\eqref{eq:BlochSSH} and obtain
\begin{equation}
    \partial_\lambda \phi_+(x) = \frac{1}{E_{s,\alpha}}\begin{pmatrix}
    0 \\
    \frac{1 - \alpha^4 - 4\lambda \alpha \partial_\lambda \alpha}{(\lambda-1)\alpha^3 - (1+\lambda) \alpha}
    \end{pmatrix} \ ,
\end{equation}
and
\begin{equation}
    \partial_\lambda \phi_-(x) = \frac{1}{E_{s,\alpha}} \begin{pmatrix}
    0 \\
    \frac{1 - \alpha^4 - 4\lambda \alpha \partial_\lambda \alpha}{(1+\lambda)\alpha^3 - (1-\lambda) \alpha}
    \end{pmatrix} \ .
\end{equation}
By differentiating the local Schr\"odinger equation of Eq.~\eqref{eq:localSE} for this example, we can also obtain for the in-gap states that
\begin{equation}
    \partial_\lambda \alpha = -\alpha \frac{1+2\alpha^2+\alpha^4+\lambda^3-2\alpha^2\lambda^3+\alpha^4\lambda^3}{\lambda\left( \alpha^4-1\right)\left( 
\lambda-1 \right)\left( 1+\lambda \right)^2} \ .
\end{equation}
Note, the bulk states of a commensurate lattice have $\partial_\lambda \alpha=0$ as discussed in Sec.~\ref{sec:CDlat}.

\begin{figure}[t]
	\centering
	\includegraphics[width=0.98\linewidth]{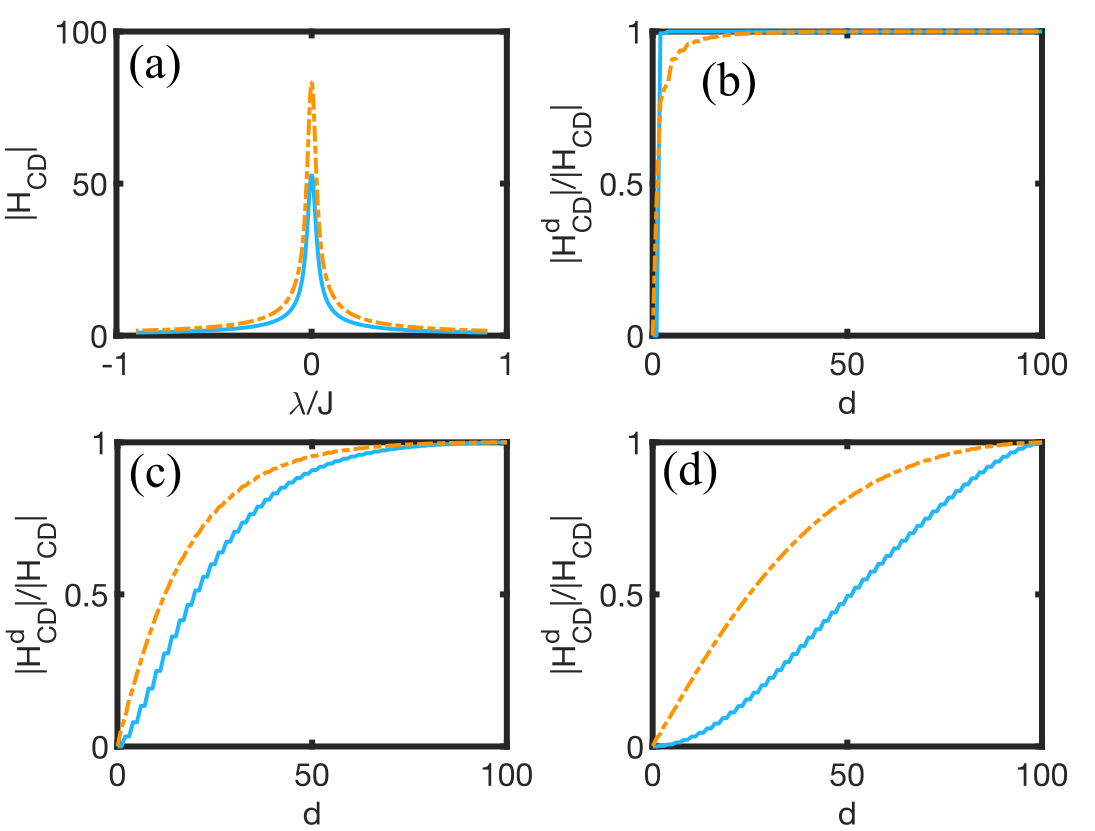}
	\caption{The norm of the CD terms in the SSH model for the same example as shown in Fig.~\ref{fig:CDSSH} for $101$ sites. Protocol is from $\lambda/J=0.9$ to $\lambda/J=-0.9$ in a total time of $\tau J = 1$. We show the case of all CD terms given by $H_\mathrm{CD}$ as dash-dot (orange) lines and the targeted CD terms given by $H_\mathrm{CD}^\alpha$ using only the in-gap state as a solid (blue) line. (a) Shows the Euclidean norm of the CD terms as a function of $\lambda/J$, showing a clear increase in the `strength' of the CD terms around the gap closing point of $\lambda/J=0$. (b-d) Shows the percentage of the Euclidean norm that is accounted for when including $d$ diagonals of the matrix at (b) $\lambda/J=0.9$, (c) $\lambda/J=0.046$, and (d) $\lambda/J=10^{-3}$, with all of $H_\mathrm{CD}$ being accounted for at $100$ diagonals.}
	\label{fig:SSHNorm}
\end{figure}

\begin{figure}[t]
	\centering
	\includegraphics[width=0.98\linewidth]{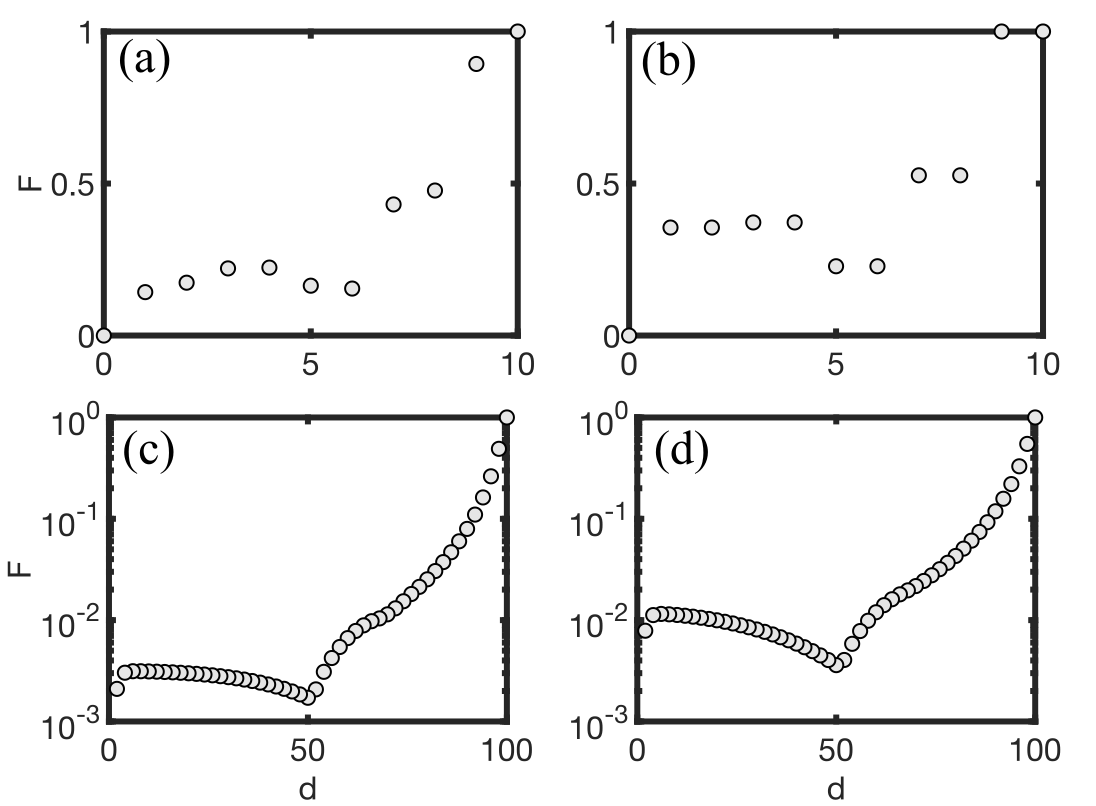}
	\caption{The state transfer fidelity after implementing CD including up to $d$ diagonals of the CD Hamiltonian for the same protocol as in Fig.~\ref{fig:SSHNorm}. (a,b) Show the case of $11$ sites and (c,d) $101$ sites with (a,c) being the implementation of all CD terms given by $H_\mathrm{CD}$ and (b,d) the targeted CD terms given by $H_\mathrm{CD}^\alpha$ using only the in-gap state. The state transfer fidelity after implementation of the bare protocol with no CD is $F=10^{-10}$ for $11$ sites and $F<10^{-14}$ for $101$ sites.}
	\label{fig:DynamicsDiag}
\end{figure}

\subsubsection{All states}

First, we will consider the normal form of CD, which includes the corrections for all instantaneous eigenstates. The form of CD for all states for non-interacting lattice models is given in Eq.~\eqref{eq:CDall}, and is fully characterised for each potential hopping term between $x$ and $x^\prime$ by
\begin{equation}
    \Theta(x,x^\prime) = \sum_\alpha \theta_\alpha(x,x^\prime) \ .
\end{equation}
We will consider the state transfer protocol in the SSH model by initialising a system at $\lambda_0=0.9$ in the in-gap state then linearly driving $\lambda$ to $\lambda_f=-0.9$ thus transferring the state from the left to the right edge of the system as is shown in Fig.~\ref{fig:SSHExample}. Imposing the evolution of the system under $H+H_\mathrm{CD}$, with $H_\mathrm{CD}$ given in Eq.~\eqref{eq:CDall}, we obtain unit fidelity state transfer across the lattice for all total driving times and system sizes.

We show the strength of the tunnelling terms between all $x$ and $x^\prime$ in the upper triangles of Fig.~\ref{fig:CDSSH}, the operator is Hermitian so symmetric up to a local phase around $x=x^\prime$. Two examples are shown for a small and large lattice with $11$ and $101$ sites respectively, and we find that the form of the CD term is very similar through various commensurate system sizes. We find that as we approach the gap-closing point, the tunnelling terms required to enforce CD become longer in range, reflecting the more delocalised nature of the state at this point, see Fig.~\ref{fig:SSHExample}(c). We repeat here that implementation of the full modified Hamiltonian $H+H_{\mathrm{CD}}$ with the form of $H_{\mathrm{CD}}$ given by the upper triangles of Fig.~\ref{fig:CDSSH} results in unit fidelity state transfer in arbitrary time.

\subsubsection{Targeted state}

We now consider the case of CD for only a single state as described by Eq.~\eqref{eq:CDparticular}, which is particularly useful in this state preparation scenario where we desire to remain in a single instantaneous eigenstate of the system. We can characterise the strength of the tunnelling  simply with $\theta_\alpha(x,x^\prime)$ with $\alpha$ corresponding to the state of interest, in this case the in-gap state. Imposing the evolution of the system under $H+H_\mathrm{CD}^\alpha$, with $H_\mathrm{CD}^\alpha$ given in Eq.~\eqref{eq:CDparticular}, we again obtain unit fidelity state transfer across the lattice for all total driving times and system sizes.

We show the strength of the tunnelling terms in order to enforce CD for the in-gap state in the lower triangles of Fig.~\ref{fig:CDSSH}, for systems of size $11$ and $101$. Again, the operator is Hermitian so symmetric up to a local phase around $x=x^\prime$. A number of differences to the full CD terms are immediately observed. First, at large $|\lambda|$ the CD terms to correct for diabatic excitations away from the in-gap state only require tuning the tunnelling coefficients near the edge that is populated. While this may be easier to realise in a given physical set-up, as it does not require tuning all tunnelling coefficients at all ranges, it creates an overall aperiodic Hamiltonian. This should not be a surprise to us, as the in-gap edge states are inherently asymmetric, as they are bound to a particular edge. When $|\lambda|$ is decreased, we observe again that longer-range tunnellings come into play. However, this does not become as uniform as in the case of the full CD with the longest tunnelling terms dominating as we approach $\lambda=0$.

\subsubsection{Properties and dynamics}

It is known that the norm of a CD term can be sensitive to points where energy gaps close \cite{berry_transitionless_2009,pandey2020adiabatic,hatomura2021controlling,lawrence2023numerical}, this is a particular issue if there is only one gap present in the system. We plot the Euclidean norm of the CD terms for both the full CD and targeted CD in Fig.~\ref{fig:SSHNorm}(a) for $101$ sites and confirm that it does diverge around the gap closing point. Note, that in any finite system, this divergence will appear finite as we will have a small but finite gap between the states.

From the forms of the CD given in Fig.~\ref{fig:CDSSH}, it is clear that a key feature is that as we decrease $\lambda$ towards zero, we populate the diagonals further away from the central diagonal of $\theta_\alpha(x,x^\prime)$ or $\Theta(x,x^\prime)$ with non-zero entries. In other words, long-range tunnellings become increasingly important. In Fig.~\ref{fig:SSHNorm}(b-d), we plot the ratio of the norm of the CD terms including up to $d$ diagonals below the central diagonal, with all other values set to zero, and the norm of the CD term for three different $\lambda$ values. The ratio of norms for the case of large $|\lambda|$ is shown in Fig.~\ref{fig:SSHNorm}(b) with both the full and targeted CD described by only a few diagonals as would be expected from the local nature of the CD tunnelling shown in Fig.~\ref{fig:CDSSH}(e). When $|\lambda|$ is decreased, as is shown in Figs.~\ref{fig:SSHNorm}(c) and (d), the CD terms become more non-local in nature and begin to be dominated by terms that are longer in range. 

The impact of the highly non-local CD terms in the final achieved state transfer fidelity is shown for both the targeted and full CD approaches in Fig.~\ref{fig:DynamicsDiag}. Across both approaches, it is clear in Fig.~\ref{fig:DynamicsDiag} that the non-local terms are crucial to the transfer of the state, as would be expected, as the gap closing at $\lambda=0$ will have a significant impact on the diabatic transitions during the protocol.

\begin{figure}[t]
	\centering
	\includegraphics[width=0.98\linewidth]{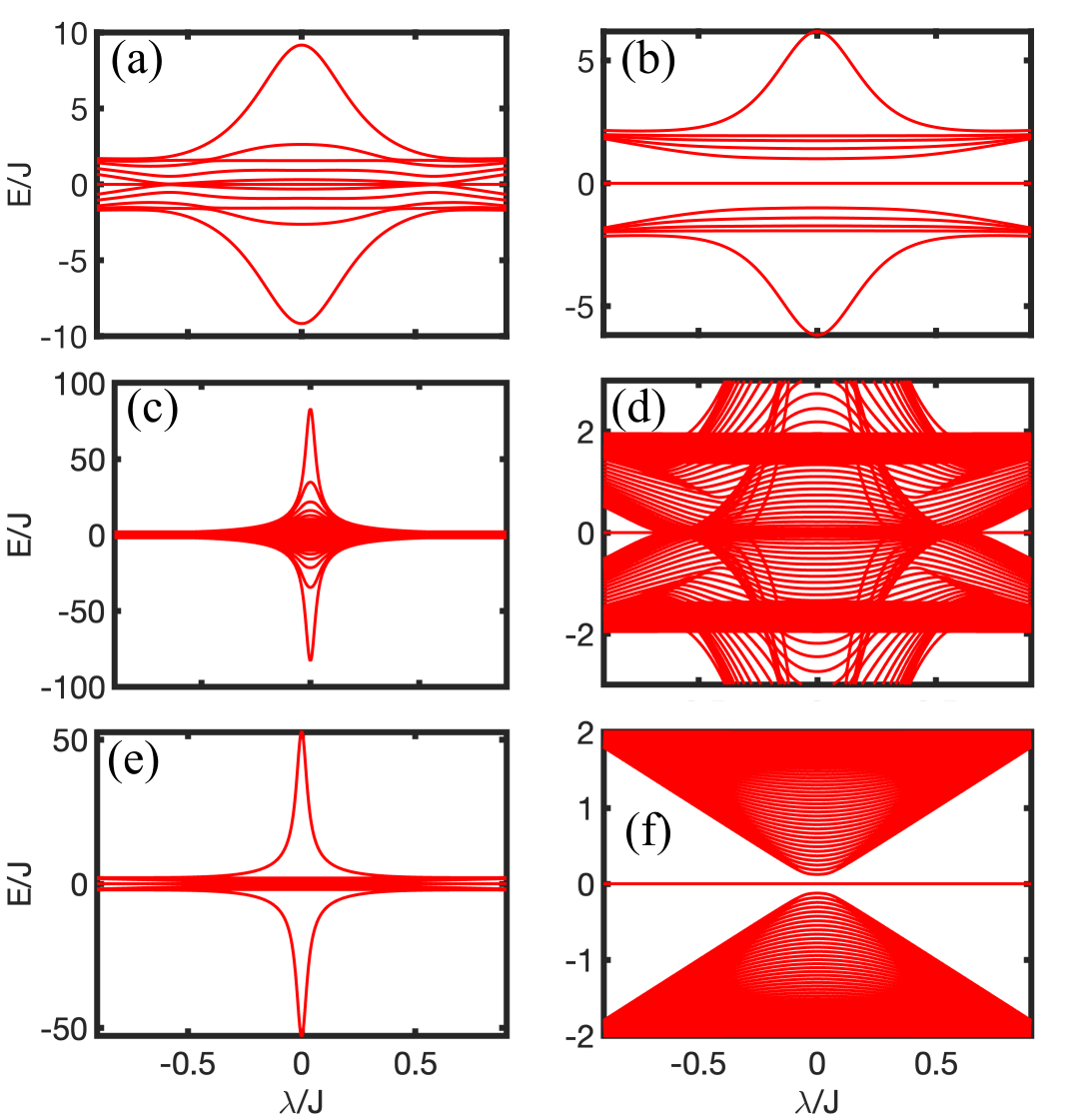}
	\caption{Examples of the spectrum of the full Hamiltonian with CD for both (a,c,d) the full CD term so the Hamiltonian is $H(\lambda)+ H_\mathrm{CD}(\lambda)$ and (b,e,f) the targeted CD term for the in-gap state so the Hamiltonian is $H(\lambda)+ H_\mathrm{CD}^\alpha(\lambda)$. Examples are shown for a total drive time of $\tau J=1$ for systems of size (a,b) $11$ sites and (c-f) $101$ sites. The plots in (d) and (f) are zoomed in portions of (c) and (e) respectively to see the spectrum around the central $E/J=0$ state. The CD terms are found using the analytical approach outlined in Sec.~\ref{sec:CDlat} and the eigenvalue problem is solved numerically.}
	\label{fig:CDspectra}
\end{figure}

\subsection{Spectra of the modified counterdiabatic Hamiltonian}

Given that we have obtained the full analytical CD terms, we can also consider the spectra of the modified Hamiltonians under which we evolve when we add CD, given by $H(\lambda) + H_\mathrm{CD}(\lambda)$. We plot these in Fig.~\ref{fig:CDspectra} for the case of $11$ and $101$ sites and the driving protocol considered so far. We solve for both the full and targeted CD using the analytical approach outlined in Sec.~\ref{sec:CDlat} but numerically solve for the instantaneous eigenstates corresponding to $H(\lambda) + H_\mathrm{CD}(\lambda)$, as the modified Hamiltonian can be aperiodic and non-sparse, due to the asymmetric form of the CD Hamiltonian, as is shown in Fig.~\ref{fig:CDSSH}. The analytical approach can still be applied in this regime but is rather inefficient in this scenario as it will require $\tau = L-x_0$ non-local coupled Schr\"odinger equations to be solved, especially around the gap closing point. 

\begin{figure}[t]
	\centering
	\includegraphics[width=0.98\linewidth]{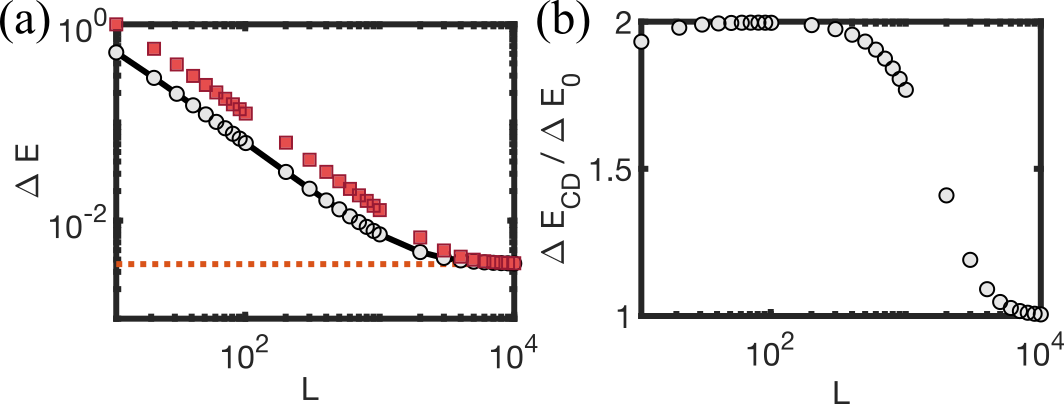}
	\caption{Scaling of the gap near to the topological transition, $\lambda/J = 1.8 \times 10^{-3}$. (a) The scaling of the energy gap between the zero energy topological edge state and the nearest state in energy. The analytical energy gap, given by Eq.~\eqref{eq:SSHgapclose}, is shown by a solid (black) line which is in agreement with the numerical gap for the SSH model with Hamiltonian~\eqref{eq:SSHH} by circles (grey). The gap for the CD Hamiltonian of the targeted approach, shown in Fig.~\ref{fig:CDspectra}(b) and (f), is given by squares (red). We also plot the limit $\lim_{L \rightarrow \infty} \Delta E (\lambda) = 2\lambda$ by a dotted (red) line. (b) The ratio of the gap of the CD Hamiltonian targeting the edge state, $\Delta E_\mathrm{CD}$, to the analytical energy gap, $\Delta E_0$, of the SSH model.}
	\label{fig:GapScale}
\end{figure}

The impact of the inclusion of CD terms targeted at a particular state is clear for the case of $11$ sites in Fig.~\ref{fig:CDspectra}(b) as the result is that an increased minimal gap has been opened between the state at energy zero and the neighbouring states, with this gap being $\Delta E/J = 0.261$ before the addition of the CD term for the in-gap state and $\Delta E/J = 1$ with the CD term. Note, that we can calculate the minimal gap without the CD term from Eq.~\eqref{eq:SSHgapclose}. When we go to a larger system of $101$ sites, Fig.~\ref{fig:CDspectra}(e,f), it does not immediately look like the minimal gap has been made larger as, of course, all of the eigenstates are far closer. However, in this case, we again see that the gap has been increased from $\Delta E/J = 0.031$ without CD to $\Delta E/J = 0.124$ with the inclusion of CD terms for the in-gap state. Note, in order to open this gap in each case the CD terms impose a penalty in energy (away from zero) for some states, resulting in states being pushed out the top and bottom of the bands around $\lambda/J=0$. For larger systems, this penalty appears more severe, and it is this that we observe as a divergence in the norm of the CD terms at the gap closing point in Fig.~\ref{fig:SSHNorm}(a). 

In Fig.~\ref{fig:GapScale} we consider the scaling of the gap for a small finite $\lambda/J=1.8 \times 10^{-3}$ for both the original Hamiltonian and the Hamiltonian with targeted CD. We observe that the gap obtained from numerical diagonalisation for the original Hamiltonian, $H(\lambda)$, is in agreement with the analytical form derived earlier, see Eq.~\eqref{eq:SSHgapclose}, and that it converges to the infinite size limit of $2\lambda$. We observe that the energy gap for the Hamiltonian with targeted CD, $H(\lambda)+ H_\mathrm{CD}^\alpha(\lambda)$, is a factor of two larger than the energy gap in the non-modified Hamiltonian, before eventually also converging to the infinite size limit value of $2\lambda$.

The spectra for the modified Hamiltonian for the full CD term are shown in Figs.~\ref{fig:CDspectra}(a) for $11$ sites and (c,d) for $101$ sites. There are similarities between these and the spectra of the targeted CD terms, mainly states being pushed out to energies far from zero, resulting in a similar behaviour of the norm. It is clear though that the spectrum of this modified Hamiltonian is far more complex and it is difficult to build an understanding of how this modified Hamiltonian imposes the adiabatic approximation in arbitrary time from its spectra alone. It is clear that the full CD term is not opening a gap for the edge state. We know this is not necessary as the eigenstates of the modified Hamiltonian are not required to be equivalent to or bear any resemblance with the eigenstates of the bare Hamiltonian with no CD. Further study of the connection between the eigenstates of the CD and bare Hamiltonians could provide insight into the key requirements for developing fast protocols for more complex settings.

\section{Conclusion and Outlook}

In this work, we have extended the set of known examples where the exact analytical form of CD can be found to include the general Hamiltonian of non-interacting particles, which can be either bosons or fermions, in 1D lattices. We built upon previous work \cite{duncan2018exact} that outlined the analytical states for both bulk and edge states to construct the CD terms exactly. We outlined how this approach can be applied to general problems in this family of Hamiltonians and derived expressions for terms that appear in the CD. We also discussed a targeted approach for countering transitions out of a particular state.

As an example, we considered the CD terms, both for all states and targeted at a particular state, for state transfer in the topological SSH model. The approach developed enabled us to study the properties of the CD terms for large systems with no penalty due to the size of the system. Note, that while the Hilbert space is only increasing linearly with system size, the analytical approach outlined can still outperform numerically obtaining the eigenstates, especially for the commensurate case. We restricted the shown results to a lattice of $101$ sites as this allowed for the behaviour of large systems to be observed clearly in the plots, and the CD terms can be obtained analytically for arbitrarily large system sizes.

The approach outlined in this work could also be extended to higher-dimensional systems with open boundaries that can be reduced to one-dimensional models, e.g., crystalline models in cylindrical geometries. It could also be extended to study few-body systems where the wave functions can be written as combinations of the non-interacting states, or the CD terms obtained here could be used as starting points for control functions in many-body systems for variational CD approaches \cite{sels2017minimizing,cepaite2023counterdiabatic}. 

The requirement to control terms over arbitrarily long distances will be a common feature of CD terms in lattices, as the CD operator is generated by a term proportional to $\ket{\partial_\lambda \psi_n(\lambda)} \bra{\psi_n(\lambda)}$. We observed this in the non-interacting case for the tunnelling terms in the SSH model, especially as we approached the point where the band gap closed. For the many-body case, it is highly likely that exact CD will require the full extended Hubbard \cite{le2017extended} or Bose-Hubbard model \cite{dutta2015non} to be controlled, i.e., long-range tunnelling, interaction, pair tunnelling, and density-dependent tunnelling terms. 
It is possible to engineer the terms of the extended Bose-Hubbard model, including long-range tunnelling, through placing ultracold atoms in optical lattices in a cavity \cite{ritsch2013cold,landig2016quantum,arguello2019analogue}. While it is possible that such an approach could in principle realise the CD terms of this work, it is unlikely that the structure of the exact CD terms discussed here could be easily engineered.

However, knowledge of the exact CD terms provides us with additional information about the dynamics of a quantum system outside of the adiabatic approximation. For example, the adiabatic gauge potential, which for a choice of gauge is equivalent to the CD Hamiltonian up to a global phase, has been utilised as a numerically efficient cost function for optimal control protocols \cite{cepaite2023counterdiabatic}, to define and probe the properties of chaotic behaviour \cite{pandey2020adiabatic,lim2024defining}, to probe the presence of quantum phase transitions \cite{hatomura2021controlling,lawrence2023numerical}, and, in general, to study non-equilibrium behaviour \cite{kolodrubetz2017geometry}. 

\acknowledgements{The author thanks Andrew J. Daley, Ewen D. C. Lawrence, Stewart Morawetz, and Ciar\'{a}n Hickey for helpful discussions. This work was supported by the Engineering and Physical Sciences Research Council through grant EP/Y005058/2.}



%

\end{document}